\documentclass{phbauth}
\usepackage{graphicx}

\begin{document}

\begin{frontmatter}

\title{Perpendicular transport properties of 
	YBa$_2$Cu$_3$O$_{7-\delta}$/PrBa$_2$Cu$_3$O$_{7-\delta}$ superlattices.}

\author[address1]{{J.C.}{Mart\'\i nez}\thanksref{thank1}},
\author[address1]{{A.}{Schattke}},
\author[address1]{{G.}{Jakob}}
\author[address1]{{H.}{Adrian}}

\address[address1]{Johannes Gutenberg - Universit{\"a}t Mainz; Institute of 
Physics; 
55099 Mainz; Germany}

\thanks[thank1]{Corresponding author. E-mail: martinez@mail.uni-mainz.de}

\begin{abstract}
The coupling between the superconducting planes of YBa$_2$Cu$_3$O$_{7-\delta}$ / 
PrBa$_2$Cu$_3$O$_{7-\delta}$ superlattices has been measured by c-axis 
transport. We show that only by changing the thickness of the superconducting 
YBa$_2$Cu$_3$O$_{7-\delta}$ layers, it is possible 
to switch between quasi-particle and Josephson tunneling.
From our data we deduce a low temperature c-axis coherence length of $\xi_c=0.27$\ nm.
\end{abstract}

\end{frontmatter}


Artificial  YBa$_2$Cu$_3$O$_{7-\delta}$/PrBa$_2$Cu$_3$O$_{7-\delta}$ 
superlattices (Y123/Pr123) constitute ideal model systems for isolating 
given properties of High Temperature Superconductors. In those systems it 
is possible for instance to modify the c-axis tunneling properties simply by 
varying the periodicities of the Y123 and Pr123 layers.

Series of 200\ nm thick Y123/Pr123 superlattices have been prepared by 
high-pressure dc-sputtering. The high quality of the samples was checked by 
detecting up to third order satellite peaks in x-ray $\theta-2\theta$ scans. 
Later on series of mesa structures with dimensions between $15\times 15$ and 
$50 \times 50$\ $\mu$m$^2$ were prepared by ion milling. In this 
work we present low temperature data on 2:7 (2 layers of Y123 and 7 of Pr123) 
and 8:8 superlattices.

\begin{figure}[ht]
\begin{center}\leavevmode
\includegraphics[width=1\linewidth]{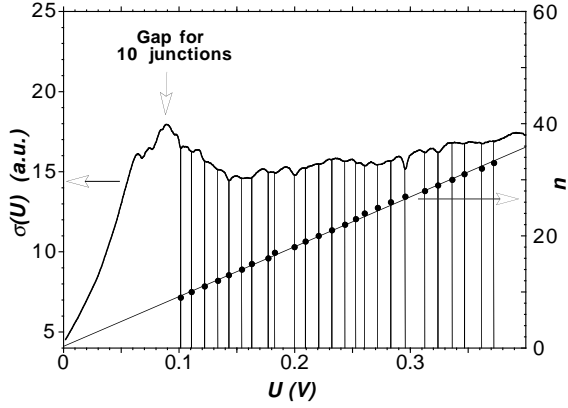}
\caption{ 
Differential conductance $\sigma$ versus Voltage $U$ for a 2:7 superlattice. The 
vertical lines mark minima observed in the spectrum. The right axis corresponds 
to the index associated to each minimum. The linear fit reveals the existence 
of a quasi-periodicity.
}\label{fig1}\end{center}\end{figure}

We measured simultaneously the $U$ vs. $I$ characteristics and the differential 
conductivity $\sigma(U)$ by means of a standard Lock-In technique. In Fig.\ 
\ref{fig1} we show the $\sigma(U)$ on a $30\times 30 \mu$m$^2$ 
mesa done on a 2:7 superlattice at 2.0\ K. 
No superconducting current could be detected. However the peak in 
$\sigma(U)$ corresponds to a $c$-axis superconducting gap. From the peak to peak voltage $U_{pp}$ 
peak, we estimate that each of the $n=$8 to 10 bi-layers constituting this mesa 
have a $c$-axis gap $\Delta_c=U_{pp}/4 n=5.0\pm 0.5$ \ meV. 
This value is in excellent agreement with the value of $\Delta_c$ given in 
the literature which scatters between 4 and 6 meV for planar junctions \cite{iguchi}. 

In Fig.\ \ref{fig1}, we observe sharper features in $\sigma(U)$. In order to 
verify a quasi-periodicity, we marked each minimum by a vertical line which is 
associated to an integer. In Fig.\ \ref{fig1} 
we plot this index as a function of the minima position. A clear zero crossing of the 
linear fit is obtained by choosing an index n=9 for the lowest index. 
From the linear fit we deduce a period of $(11.1 \pm 0.5)$\ mV which gives 
a periodicity $\delta U=(1.2\pm 0.1)$ meV for a single junction. These 
features are reproducible and temperature independent up to $20$\ K.

Such a quasi-periodic structure in the density of states has been theoretically 
predicted by Hahn \cite{hahn}. According to this work, above the superconducting 
gap, additional structures should appear with a periodicity of: 
$\xi_c/s=1/\pi^2 \delta U/\Delta_c$ where $s$ is the period of the superlattice and 
$\xi_c$ the $c$-axis coherence length. From our data 
and by taking $s=10.5$\ nm we deduce a $c$-axis coherence length of $\xi_c=0.27$ 
nm. If we assume for Y123 an anisotropy of $\gamma \approx 5$ \cite{janossi} we 
would obtain a in-plane coherence length $\xi_{ab}=\gamma\xi_c\approx 1.4$\ nm. 
This value is close to the generally quoted $\xi_{ab}=1.5$\ nm for Y123 
\cite{tinkham}.

\begin{figure}[h]
\begin{center}\leavevmode
\includegraphics[width=1\linewidth]{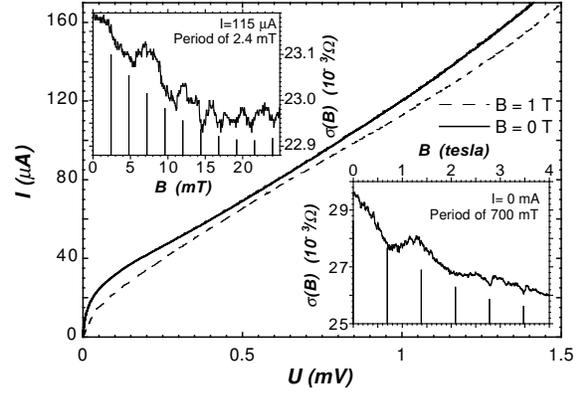}
\caption{ 
Differential conductance $\sigma$ versus Voltage $U$ for a 8:8 superlattice. The 
inserts correspond to the field dependence of $\sigma(U=0)$ at two different 
current bias.
}\label{fig2}\end{center}\end{figure}

In Fig.\ \ref{fig2}, we show the $I$ vs. $U$ characteristics of a $40\times 40 \mu$m$^2$ 
mesa done on a 8:8 
superlattice for zero field and for $B=1$\ T. From the difference between the 
two curves we deduce the presence of two distinct low and high current regimes. 
To investigate the difference between these two regimes we measured the $B$ 
dependence of $\sigma(U=0)$ at zero bias current and at $115$\ $\mu$A (see 
inserts).

At zero current $\sigma(B)$ shows a modulation of $B_\Phi=0.7$\ T. By 
considering that $B_\Phi=\Phi_0/(s+t)b_{eff}$, s=9.4 nm (Y123 thickness) and 
t=9.4 nm (Pr123 thickness), we deduce $b_{eff}=0.152$\ $\mu$m. The low current 
regime can therefore be associated to structural shorts \cite{remark2}. At $115$\ $\mu$A we 
observe a $\sigma(B)$ modulation of $2.4$\ mT. This value is very similar to the 
$2.5$\ mT predicted for a $40\times 40$\ $\mu$m$^2$ mesa. The high current 
regime can therefore be associated to a c-axis Josephson Effect. The absence of a 
similar effect in the 2:7 superlattices is probably due to a non-fully developed 
superconducting order parameter in the 2 unit cells thick Y123 layers.

This work was supported by the German BMBF (Contract 13N6916) and the E.U. 
(ERBFMBICT972217).


\begin{thebibliography}{9}
\bibitem{remark2} Such shorts can mainly occur in superlattices having similar thicknesses for both layers. Superlattices having Y123 thicknesses much smaller than the Pr123 layer show usually no shorts.
\end{thebibliography}
\end{document}